\documentclass[12pt,preprint,apj]{emulateapj}
\newcommand{\kmsMpc}{{km s$^{-1}$ Mpc$^{-1}$}}

\usepackage{epstopdf}
\usepackage{color}

\shorttitle{ TRGB Distances to NGC 3021, NGC 3370 and NGC 1309}
\shortauthors{Jang \& Lee}

\begin{document}

\title{
% {\color{red} { \bf Revised DRAFT  \today} } \\
%{{ \bf To be Submitted} } \\
THE TIP OF THE RED GIANT BRANCH DISTANCES TO TYPE IA SUPERNOVA HOST GALAXIES. V. \\
NGC 3021, NGC 3370, and NGC 1309 and THE VALUE OF THE HUBBLE CONSTANT
}
 
\author{In Sung Jang and Myung Gyoon Lee }
\affil{Astronomy Program, Department of Physics and Astronomy, Seoul National University, Gwanak-gu, Seoul 151-742, Korea}
\email{isjang@astro.snu.ac.kr, mglee@astro.snu.ac.kr }

%==============================================================================================================

\begin{abstract}
We present final results of a program for the determination of the Hubble constant based on the calibration of the Type Ia supernovae (SNe Ia) using the Tip of the Red Giant Branch (TRGB).
We report TRGB distances to three SN Ia host galaxies, NGC 3021, NGC 3370, and NGC 1309. 
We obtain F555W and F814W photometry of resolved stars 
from the archival $Hubble$ $Space$ $Telescope$ data. 
Luminosity functions of red giant stars in the outer regions of these galaxies show the TRGB to be at $I \approx QT=28.2\sim28.5$ mag.
From these TRGB magnitudes and the revised TRGB calibration based on two distance anchors (NGC 4258 and the LMC) in \citet{jan16}, we derive the distances: $(m-M)_0=32.178\pm0.033$ for NGC 3021, $32.253\pm0.041$ for NGC 3370, and $32.471\pm0.040$ for NGC 1309.
We update our previous results on the TRGB distances to five SN Ia host galaxies using the revised TRGB calibration. 
By combining the TRGB distance estimates to SN Ia host galaxies  in this study with the SN Ia calibration provided by \citet{rie11}, we obtain a value of the Hubble constant: $H_0=71.66\pm1.80 (\rm random)\pm1.88 (\rm systematic)$ \kmsMpc~ (a 3.6\% uncertainty including systematics) from all eight SNe, and $H_0=73.72\pm2.03\pm1.94$ \kmsMpc~ (a 3.8\% uncertainty) 
from six low-reddened SNe.
We present our best estimate, $H_0=71.17\pm1.66\pm1.87$ \kmsMpc~ (a 3.5\% uncertainty)  
from six low-reddened SNe with the recent SN Ia calibration in \citet{rie16}.
This value is between those from the Cepheid calibrated SNe Ia  and those from the Cosmic Microwave Background (CMB) analysis,
lowering the Hubble tension.

\end{abstract}


\section{Introduction}

Measuring the value of the Hubble constant ($H_0$) is a fundamental %and a critical 
step in extragalactic astronomy and cosmology.
Tremendous work has been conducted over the past few decades %in order
 to accurately determine the value of the Hubble constant. 
%A tremendous work to determine the value of the Hubble constant has been conducted over the past decades. 
However, the estimated values have been controversial until recently. 
In the 20th century, two groups led by Allan Sandage and Gerard de Vaucouleurs, respectively,  measured the value of the Hubble constant to be between $H_0$= 50 and 100 \kmsMpc \citep{dev81, dev93, san74a, san74b, san74c, san74d, san75a, san75b, san76, van60a, van60b, van75, van94}.
%In the 20th century, there was a strong controversy of $H_0$ between $H_0$= 50 and 100 \kmsMpc \citep{dev81, dev93, san74a, san74b, san74c, san74d, san75a, san75b, san76, van60a, van60b, van75, van94}, by the two groups led by Allan Sandage and Gerard de Vaucouleurs.
This was known as "the factor of two controversy".
%It is called "the factor of two controversy".
Although it is not clear that this large discrepancy of $H_0$ originated from a purely scientific result or from a publication bias, the value of $H_0$ has settled down to a value between the two, $H_0\sim70$ \kmsMpc, with the advent of the $Hubble$ $Space$ $Telescope$ (HST). 
%Although it is not clear that this large discrepancy of $H_0$ is originated from a pure scientific result or a publication bias, the value of $H_0$ has settled down between the two, $H_0\sim70$ \kmsMpc, with the advent of the $Hubble$ $Space$ $Telescope$ (HST). 

The Hubble Key Project played a key role in the calibration of $H_0$ \citep{fre01}. They improved the luminosity calibration of Type Ia Supernovae (SNe Ia) with the HST and yielded a value of $H_0$ = $71\pm2 (\rm random)\pm6 (systematic)$ \kmsMpc. % \citep{fre01}. 
A few years later, however, the SN HST project presented a somewhat lower value, $H_0=62.3\pm1.3 (\rm random)\pm5.0 (\rm systematic)$ \kmsMpc~ \citep{san06} from nearly the same HST imaging data as used in the Hubble Key Project. The level of difference is $\sim1.5\sigma$.
%, which is a weak tension. 
Recent estimations by the Supernovae and H0 for the Equation of State (SH0ES) and the Carnegie Hubble Program provided more precise values: $H_0=73.8\pm2.4$ \kmsMpc ~ \citep{rie11}, $H_0=73.02\pm1.79$ \kmsMpc ~ \citep{rie16}, and $H_0=74.3\pm2.1$ \kmsMpc ~\citep{fre12}, respectively.
These three estimations that are based on Cepheid calibrated SNe Ia agree remarkably well.

%%%%%%%%%%%%%%%%%%%%%%%%%%%%%%%%%%%%%%%%%%%
%% TABLE 1
%%%%%%%%%%%%%%%%%%%%%%%%%%%%%%%%%
\begin{deluxetable*}{lccccccc}
\tabletypesize{\footnotesize}
%\tabletypesize{\scriptsize}
%\tabletypesize{\tiny}
\setlength{\tabcolsep}{0.05in}
%\rotate
\tablecaption{A Summary of $HST$ Observations of SNe Ia Calibration Sample}
\tablewidth{0pt}
\tablehead{ \colhead{Target} & \colhead{R.A.} &  \colhead{Decl.} & \colhead{Instrument} & \multicolumn{2}{c}{Exposure time} & \colhead{Prop. ID} \\
& (J2000.0) & (J2000.0) & & F555W  & F814W
}
\startdata
NGC 3021 	& 09 51 00.71 &  33 33 23.9 & ACS/WFC & 61,760s & 24,000s & 10497 \\
NGC 3370 	& 10 47 06.91 &  17 16 40.0 & ACS/WFC & 61,240s & 24,000s & 9651 \\
NGC 1309 	& 03 22 05.31 &--15 23 47.8 & ACS/WFC & 61,760s & 24,000s & 10497 \\
\hline
\enddata
\label{tab_obs}
\end{deluxetable*}

\begin{figure*}
\centering
\includegraphics[scale=0.8]{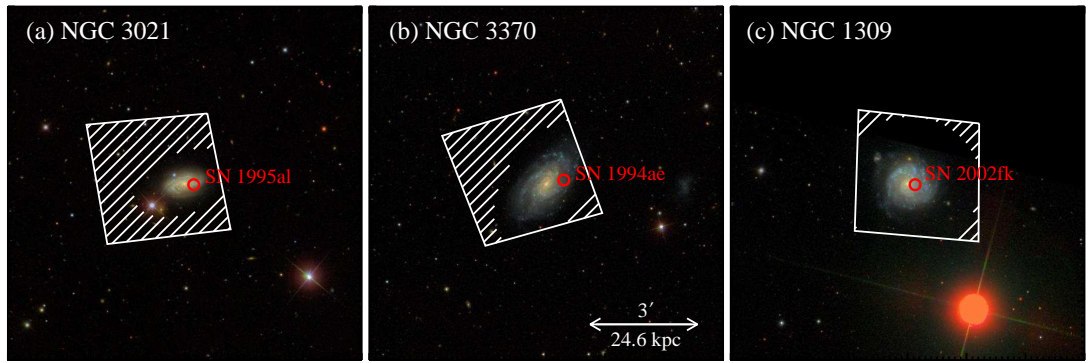} %white.eps}
\caption{Identification of HST/ACS fields of NGC 3021, NGC 3370, and NGC 1309 used in this study overlayed on the $10\arcmin\times10\arcmin$ color maps provided by the Sloan Digital Sky Survey.
Locations of the SNe Ia are marked by open circles.
Hatched regions in each HST field represent the halo regions used for the TRGB analysis.
%{\color{blue}\bf Remove RAs and Decs in Figure.}
}
\label{fig_finder}
\end{figure*}

However, the value of $H_0$ is still considered to be controversial. 
Recent analysis of the cosmic microwave background radiation (CMBR) with a flat $\Lambda$CDM cosmology by WMAP and PLANCK groups yielded values of $H_0$ with remarkably small uncertainties: $H_0=69.3\pm0.8$ \kmsMpc~ from WMAP9 \citep{ben13} and $H_0=67.8\pm0.9$ \kmsMpc~~ from PLANCK  \citep{pla15},
which are lower than the values based on Cepheid calibrated SNe. % 67.3\pm1.2 \citep{pla14}
Moreover, recent studies of the baryon acoustic oscillation (BAO) combined with SNe Ia data yielded a value  similar to those from the CMBR analysis, $H_0=67.3\pm1.1$ \kmsMpc ~ \citep{aub15}. 
These three estimations %based on the inverse distance ladder method 
agree well. However, these values are $2\sim3\sigma$ smaller than those from Cepheid calibrated SNe Ia.
Note that CMBR and BAO are inverse distance ladders, while Cepheids and SN Ia are classical distance ladders. 
This discrepancy between the results based on the inverse distance ladder method and the classical distance latter method is now one of the critical issues in modern cosmology. 
It is needed to check systematics of both the classical and inverse distance ladder methods in order to understand what may cause this discrepancy.
%An independent and precise determination of the Hubble constant would provide a critical insight over the controversy of $H_0$.

We have been working to improve the measurement of $H_0$ by measuring the accurate luminosity of SN Ia based on the Tip of the Red Giant Branch (TRGB),  as part of the program,
the TRGB distances to SN  host galaxies in the Universe (TIPSNU). 
The TRGB is known as a precise population II distance indicator \citep{lee93,fre10,jan16,bea16}.
\citet{lee12, lee13} (Papers I and II) determined the TRGB distances to three SNe Ia host galaxies: M101, M66, and M96 hosting SN 2011fe, SN 1989B, and SN 1998bu, respectively. In Paper III, \citet{jan15} derived the TRGB distances
to two additional SN Ia host galaxies: NGC 4038/39 and NGC 5584 hosting SN 2007sr and SN 2007af, respectively. By combining the TRGB distances to five SN Ia host galaxies with optical light curves for five SNe Ia in these galaxies, we obtained a value of the Hubble constant, $H_0 = 69.8 \pm 2.6 {\rm (random)} \pm 3.9 {\rm (systematic)} $ \kmsMpc. 
%Recently,
In Paper VI,  %Jang \& Lee (2016) (Paper IV) 
\citet{jan16} presented a revised calibration of the TRGB with a systematic uncertainty of 0.046 mag, much smaller than those of the previous calibrations ($\sim 0.12$ mag). This calibration used, as a distance anchor, NGC 4248 and the LMC to which geometric distances are known with high precision \citep{hum13,rie16,pie13}.

 In this paper, we present the TRGB distances to three additional galaxies hosting three SNe Ia: NGC 3021 (SA(rs)bc:) hosting SN 1995al, NGC 3370 (SA(s)c) hosting SN 1994ae, and NGC 1309 (SA(s)bc:) hosting SN 2002fk. 
Spectroscopic observations confirmed that three SNe are normal SNe Ia \citep{rie09a}.
Light curves of these SNe Ia were obtained  with modern CCD detectors,
covering their maximum luminosity epochs.
Moreover, the internal extinctions for these SNe are estimated to be small ($A_V\lesssim0.5$ mag) \citep{rie09a}. 
Thus, these three SNe Ia are excellent targets for the luminosity calibration of SN Ia.
\citet{rie05,rie09b} carried out deep HST observations for these galaxies to derive Cepheid distances to these galaxies. By combining optical light curves of Cepheid variables with their F160W band luminosities, they determined distances of 
$(m-M)_0=32.498\pm0.091$ for NGC 3021, $(m-M)_0=32.071\pm0.050$ for NGC 3370, and $(m-M)_0=32.524\pm0.056$ for NGC 1309 \citep{rie16}.
However, there are no published studies for the TRGB distances to these galaxies.

The outline of this paper is as follows.
In Section 2 we describe the data and the data reduction method.
Section 3 presents the color-magnitude diagrams (CMDs) and luminosity functions of the resolved stars in three galaxies. These will be used to estimate the TRGB distances.
In \S 4 we present updated TRGB distances to five SN Ia host galaxies included in our previous studies and compare our results with those from the SH0ES project \citep{rie11, rie16}. We also derive the absolute magnitude of SNe and the value of the Hubble constant.
Primary results are summarized in the final section.

%This paper is composed as follows. 
%Section 2 discri
%In Section 2, we 
%\S

\begin{figure*}
\centering
\includegraphics[scale=0.8]{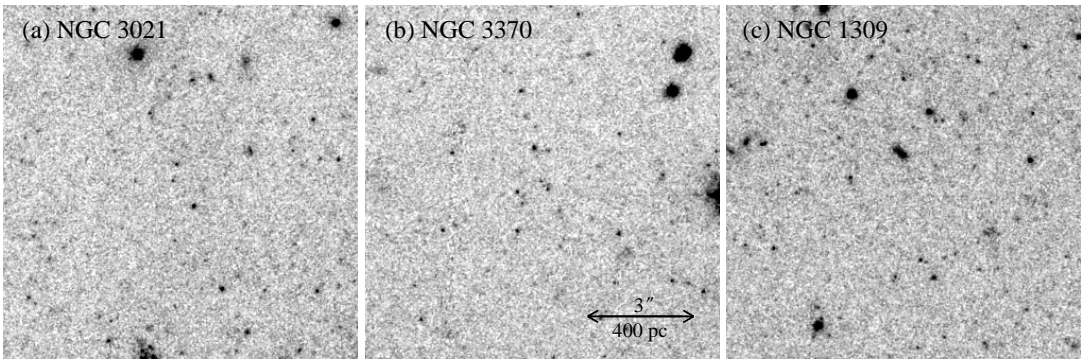} %white.eps}
\caption{
$10\arcsec \times 10\arcsec$ thumbnails of $F814W$ band drizzled images showing  the halo regions of NGC 3021 (a), NGC 3370 (b), and NGC 1309 (c). Note that most of the resolved point sources are old red giants in each galaxy.
}
\label{fig_resolve}
\end{figure*}

\begin{figure*}
\centering
\includegraphics[scale=0.9]{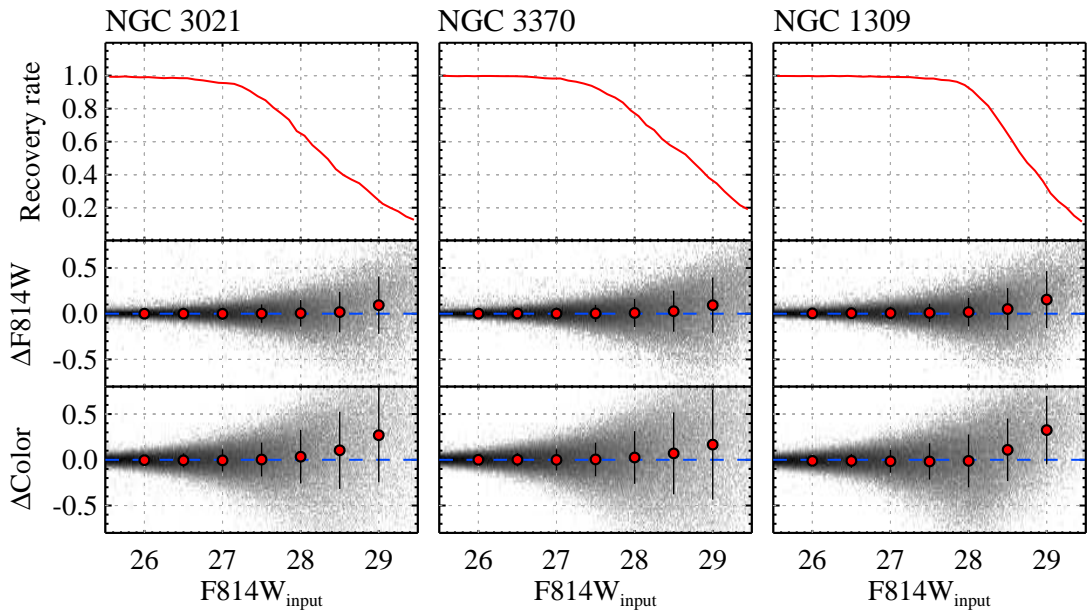} %white.eps}
\caption{ 
(Top panels) Recovery rates for RGB candidates (F555W--F814W = 1.8) in NGC 3021 (left), NGC 3370 (middle) and NGC 1309 (right).
(Middle and bottom panels) Differences in F814W and F555W--F814W (inputs minus outputs) as a function of input F814W magnitudes. Median offsets and standard deviations in each magnitude bin are marked by circles with vertical lines.
}
\label{fig_recovery}
\end{figure*}

\section{Data and Data Reduction}

{\color{red}{\bf Table \ref{tab_obs}}} lists basic information of archival HST images of the three SN Ia host galaxies we used for the TRGB analysis. 
These galaxies were observed using the same instrument, the Advanced Camera for Surveys (ACS), with a primary aim to detect Cepheid variables (PID=9651, 10497). 
These ACS fields cover halos as well as bulges and disks of the target galaxies, as shown in {\color{red}{\bf Figure \ref{fig_finder}}}.
We constructed master drizzled images of each field from individual images as described in \citet{jan15}. 
We set final\_pixfrac=0.7 and final\_scale=0$\farcs$03/pixel in the drizzling, %so that output images are expected to show 
to secure %slightly 
better angular resolution than those of drizzled images from the default setting (final\_pixfrac=1.0 and final\_scale=0$\farcs$05/pixel). Total exposure times of each field are $\sim$ 61,500s for F555W band and 24,000s for F814W band.
{\color{red}{\bf Figure \ref{fig_resolve}}} shows $10\arcsec \times 10\arcsec$ sections of combined F814W band images we made for an outer region in each galaxy. Resolved stars are clearly seen in each field, and most of the faint resolved stars are old red giant stars in the target galaxies.

Instrumental magnitudes were derived from the standard sequence, FIND--PHOT--ALLSTAR--ALLFRAME in the DAOPHOT package \citep{ste87}. PSF images were constructed using $30\sim60$ bright isolated stars in the images. 
Standard calibration of instrumental magnitudes on the HST Vega magnitude system was done using the information provided by the STScI.  Detailed reduction methods can be found in our previous papers \citep{jan15,jan16}.

We carried out artificial star experiments to check the completeness and the systematic bias of our photometry.
We added artificial stars to the selected regions of HST images that are indicated by hatched regions in {\color{red}{\bf Figure \ref{fig_finder}}} using the DAOPHOT/ADDSTAR task. 
The number of added artificial stars %added 
are $\sim$2000 for NGC 3021 and NGC 3370, and $\sim$400 for NGC 1309.
These numbers are less than 10\% of the total number of stars detected in the same regions.
We set F555W--F814W = 1.8 for the input artificial stars to make them similar to the mean color of the blue (metal poor) TRGB stars. We then performed photometry as done for real stars and derived recovery rates and photometric offsets. 
We iterated this process 100 times for NGC 3021 and NGC 3370, and 500 times for NGC 1309 to reduce statistical uncertainties. The total number of artificial stars used for the experiments is $\sim$200,000 for each galaxy.

%\clearpage

\begin{figure*}
\centering
\includegraphics[scale=0.8]{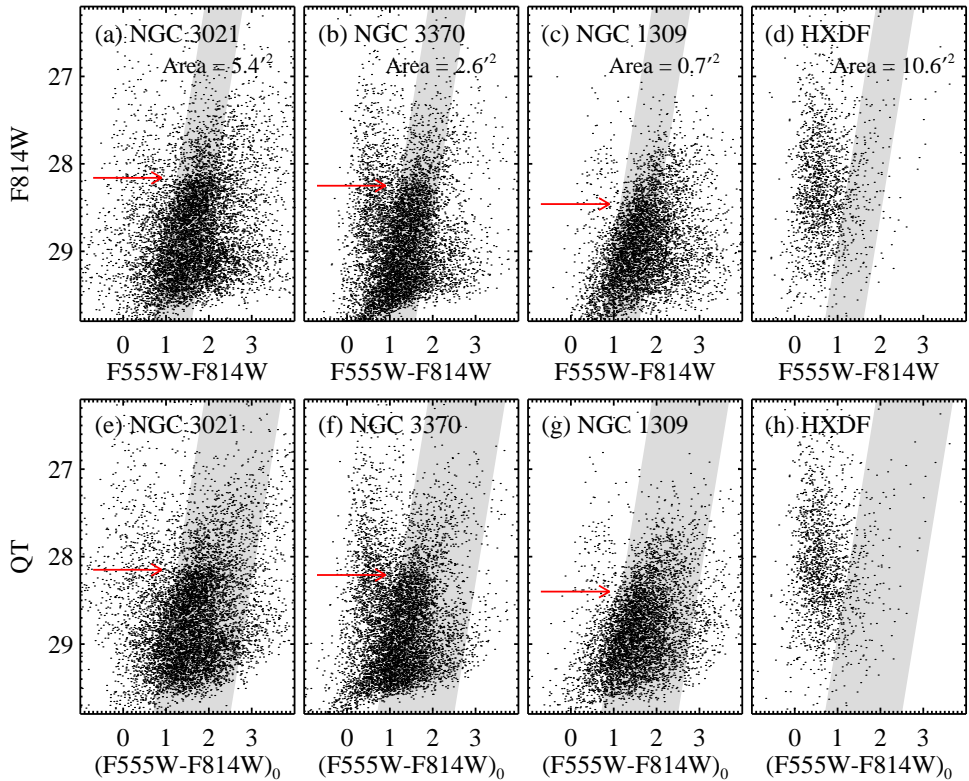} %white.eps}
\caption{(a)-(c) $F814W - (F555W-F814W)$ CMDs of the resolved stars in the halo regions of NGC 3021 (a), NGC 3370 (b), and NGC 1309 (c). We also display a CMD of the point sources in the Hubble eXtreme Deep Field (HXDF) in (d) as a reference.
(e)-(h) Same as (a)-(d), except for the $QT$ magnitude with the extinction corrected color, $(F555W-F814W)_0$. 
The shaded region in each panel denotes the boundary to select the RGB stars for the TRGB analysis. 
Estimated TRGB magnitudes are marked by red arrows.
}
\label{fig_cmd1}
\end{figure*}

%The results of the esperiments are shown in {\bf Figure \ref{fig_recovery}}. 
{\color{red}{\bf Figure \ref{fig_recovery}}} displays the result of artificial star experiments for the three galaxies.
%{\bf Figure \ref{fig_recovery}} displays the recovery rates (top panels) and the F814W magnitudes and F555W-F814W color offsets as  a function of the input F814W magnitudes.
The 50\% recovery rates are estimated to be F814W $\approx$ 28.35, 28.70, and 28.71 mag for NGC 3021, NGC 3370, and NGC 1309, respectively. 
At the TRGB magnitudes expected from the Cepheid distance estimates in the previous studies \citep{rie11, rie16} (F814W$_{TRGB} = 28.1 \sim 28.5$ mag), the recovery rates are ranging from  $\sim50\%$ (NGC 3021) to $\sim 70\%$ (NGC 3370), somewhat higher than 50\%. 
%At the TRGB magnitudes expected from the Cepheid distance estimates ($F814W_{TRGB} = 28.1 \sim 28.6$ mag) \citep{rie11, rie16}, recovery rates are ranging from $\sim50\%$ (NGC 3021) to $\sim 70\%$ (NGC 3370), somewhat higher than 50\%. 
We detect %a 
small magnitude and color offsets.
At the expected TRGB level, the F814W magnitudes and F555W--F814W color offsets are measured to be $\Delta F814W_{TRGB}\sim0.02$ mag and $\Delta(F555W-F814W)_{TRGB}\sim0.05$ mag for both NGC 3021 and NGC 3370 reductions.
The offsets are slightly larger in the case of NGC 1309 reduction: $\Delta F814W_{TRGB}\sim0.05$ mag and $\Delta(F555W-F814W)_{TRGB}\sim0.10$ mag.
Thus, measured TRGB magnitudes and colors are expected to be slightly brighter and bluer than their intrinsic values.
We considered these photometric offsets in the TRGB distance estimation (Section 3.2).

%F814W magnitude offsets (inputs minus outputs) at the TRGB magnitude are $\sim0.02$ mag for NGC 3021 and NGC 3370, and $\sim0.05$ mag for NGC 1309. Similarly, F555W--F814W mcolor offsets are $\sim0.05$ mag for NGC 3021 and NGC 3370, and $\sim0.1$ mag for NGC 1309.

%At the expected TRGB magnitudes of three galaxies (F814W$_{TRGB}$ = 28.1 $\sim$ 28.5) from the Cepheid distance estimates \citep{rie11, rie16}, recovery rates are ranging from $\sim50\%$ (NGC 3021) to $\sim 70\%$ (NGC 3370), somewhat higher than 50\%. 
%We detect a small magnitude and color offsets.

%		TRGB	delI	delCol	TRGB	deoQT	deoCol
%N3021	28.179	0.010	0.060	28.146	0.021	0.057
%N3370	28.252	0.018	0.048	28.212	0.030	0.049
%M1309	28.457	0.048	0.100	28.398	0.062	0.104

\section{Results}

\subsection{CMDs of Resolved Stars} % of the Resolved Stars

%The $HST/ACS$ fields used in this study cover bulge and disk, as well as halo region of each galaxy. 
%Sampling old RGB stars from the photometry is one of important steps in the TRGB analysis. 
%Because a mixture of younger stellar populations  
%(e.g. young main sequence stars, intermediate ages asymptotic giant branch stars) 
%enables to increase the TRGB detection uncertainties. 
The HST fields for the target galaxies cover not only the disk but also the halo of each galaxy.  
In order to sample as many old RGB stars, we selected resolved stars in the outer regions of the target galaxies, as shown by the hatched regions in {\color{red}{\bf Figure  \ref{fig_finder}}}.
%We plotted the CMDs of the resolved stars in the selected regions of the target galaxies in {\bf Figure \ref{fig_cmd1}}.  %($SMA > 1\farcm5$ for NGC 3021, $2\farcm0$ for NGC 3370, and $2\farcm1$ for NGC 1309)
%We plotted the $F814W-(F555W-F814W)$ Color Magnitude Diagrams (CMDs) resolved stars in the halo regions of NGC 3021 (a), NGC 3370 (b), and NGC 1309 (c) in Figure \ref{fig_cmd1}.
{\color{red}{\bf Figure \ref{fig_cmd1}(a-c)}} show F814W -- (F555W -- F814W) CMDs of the resolved stars in the outer regions of NGC 3021 (a), NGC 3370 (b), and NGC 1309 (c). 
All three CMDs show a prominent RGB population.
%{\color{red}{We designed a shaded region that covers  RGB stars in the CMDs.
%This region selects blue RGB stars with $1.2 \lesssim$ F555W--F814W $\lesssim 2.1$ at the TRGB level.
%}}

It is expected that some of the point sources detected in the images of the target galaxies are unresolved background galaxies. 
To estimate the background galaxy contamination in the CMDs, we investigated the CMD of the Hubble eXtreme Deep Field (HXDF), which is dominated by% background 
distant galaxies,
as shown in \citet{lee16}.
% are dominated. 
We used combined F606W and F814W images for the HXDF provided by the XDF project \citep{ill13}. 
%Total exposure times are 174,100s and 50,800s for F606W and $F814W$, respectively. 
Then we carried out PSF photometry using the same procedure as done for the target galaxies. % and constructed the CMD. 
We applied the color transformation from F606W -- F814W to F555W -- F814W using the transformation described in \citet{jan16}, to be consistent with the CMDs of the target galaxies. 

{\color{red}{\bf Figure \ref{fig_cmd1}(d)}} represents the CMD of the point sources in %the entire field of 
the HXDF. 
This CMD shows a significant vertical structure in the blue side of the shaded region (the RGB
 of the target galaxies), 
 while %it shows 
only a small number of sources are seen in the shaded region. This vertical structure represents mainly blue faint background galaxies. The area of the HXDF is 10.6 arcmin$^2$, much larger than those of the selected regions for the target galaxies (5.4 arcmin$^2$, 2.6 arcmin$^2$, and 0.7  arcmin$^2$ for NGC 3021, NGC 3370, and NGC 1309, respectively).
We counted the numbers of bright point sources with F814W $\leq29$ mag located in the shaded regions, which were chosen to select the RGB population, obtaining 2396, 1991, 1914, and 190 sources for NGC 3021, NGC 3370, NGC 1309, and the HXDF, respectively.
Considering the field area ratios, we estimated the fractions of the background galaxies for the target galaxies: 4.0\% for NGC 3021, 2.3\% for NGC 3370, and 0.6\% for NGC 1309.
Thus, background galaxy contamination in the shaded regions of the target galaxies is negligible.
We ignored the background galaxy contamination in the following TRGB analysis.
%On the contrary, the CMD of the HXDF shows an evident vertical feature at the blue color range ($1\lesssim F555W-F814W\lesssim2$). We expect that significant fractions of the blue bright sources ($F814W\lesssim29$ and $1\lesssim F555W-F814W\lesssim2$) in the CMDs of three galaxies are unresolved background galaxies.
%

\begin{figure*}
\centering
\includegraphics[scale=0.9]{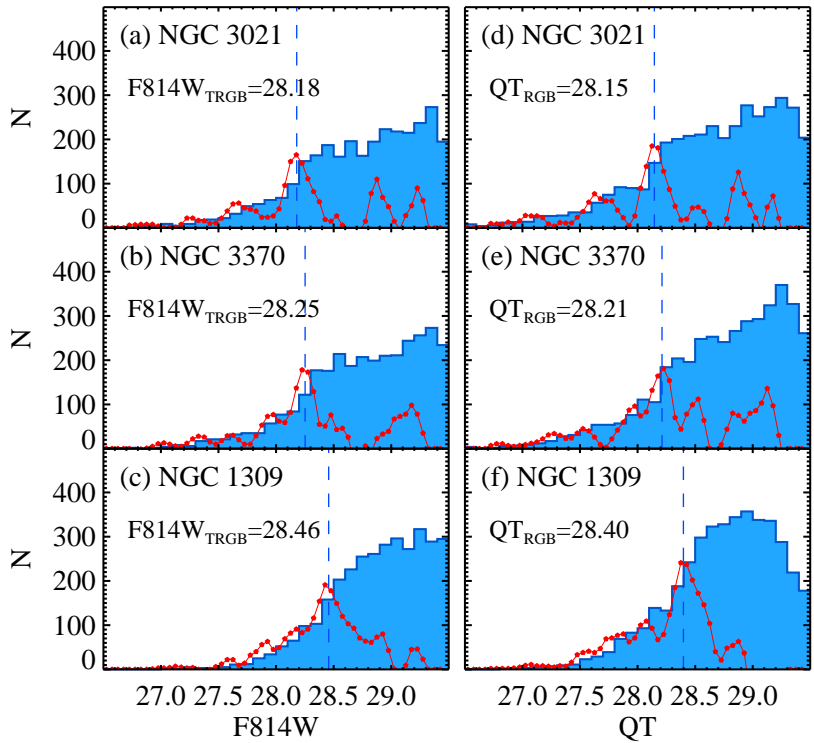} %white.eps}
\caption{(a)-(c) LFs (histograms) and edge detection responses (curved lines) of the RGB stars in NGC 3021 (a), NGC 3370 (b), and NGC 1309 (c). 
Estimated TRGB magnitudes are marked by vertical dashed lines in each panel.
(d)-(f) Same as (a)-(c), except for the $QT$ magnitude.
}
\label{fig_lf1}
\end{figure*}

\citet{jan16} introduced the $QT$ magnitude, a color (metallicity) corrected RGB magnitude, for a more accurate TRGB calibration. 
The $QT$ magnitude is defined as $QT=$ F814W$_0 - 0.116(Color-1.6)^2 + 0.043(Color-1.6)$, where $Color=($F555W -- F814W)$_0$ for the ACS/WFC.
We plotted $QT$ magnitude versus (F555W--F814W)$_0$ diagrams of the resolved point sources in {\color{red}{{\bf Figure \ref{fig_cmd1}(e-h)}}}.
Foreground reddenings are known to be very small: $E(B-V)=0.012$, 0.028, 0.035, and $0.007$ for NGC 3021, NGC 3370, NGC 1309, and the HXDF, respectively %
\citep[NASA/IPAC Extragalactic Database]
{sch11}. % 0.007 for HXDF
Internal reddenings of the selected outer regions %of the target galaxies 
are expected to be negligible so that these are assumed to be zero in this study.

\subsection{Distance Estimation} 

We determined the distances to the target galaxies using the TRGB method \citep{lee93,mad95}. 
We selected resolved stars in the shaded regions of {\color{red}{\bf Figure \ref{fig_cmd1}}}.
%, avoiding unresolved background galaxies and sampling old RGB stars as much as possible.
The shaded regions for F814W -- (F555W -- F814W) CMDs were designed to sample blue RGB stars with $1.2 \le F555W-F814W \leq 2.1$ near the TRGB level because the color (metallicity) dependence of the TRGB at the blue side is estimated to be small \citep{jan16}. In the cases of $QT$ -- (F555W -- F814W)$_0$ CMDs, 
%, which are corrected for the color dependence of the TRGB, 
we set slightly wider shaded regions, sampling RGB stars with $1.2 \le F555W-F814W \leq 3.0$ near the TRGB level.

{\color{red}{\bf Figure \ref{fig_lf1}}} displays F814W (a, b, and c) and $QT$ (d, e, and f) luminosity functions of the selected RGB stars in the CMDs. %in the outer regions of the target galaxies, which were selected using the shaded regions in the CMDs.
%Only the stars in the shaded regions of {\bf Figure \ref{fig_cmd1}} were used for the TRGB analysis.
The luminosity functions for all three galaxies show abrupt increments at $F814W\simeq QT=28.2\sim28.5$ mag, which correspond to the TRGB.
We applied an edge detection algorithm for the quantitative TRGB detection.
We used a Sobel filter employing the zero-sum kernel of [-1, -2, -1, 0, +1, +2, +1] with a bin size of 0.05 mag.
%We used a weighted logarithmic edge detector \citep{men02} described as $E(I)=\sqrt{\phi}[log[\phi(I+\sigma_I)]-log[\phi(I-\sigma_I)]]$, where $\phi(I)$ is a Gaussian-smoothed luminosity function 
%and $\sigma_I$ is the mean photometric uncertainty.
Output edge detection responses are shown as red lines in {\color{red}{\bf Figure \ref{fig_lf1}}}.

%%%%%%%%%%%%%%%%%%%%%%%%%%%%%%%%%%%%%%%%%%%
%% TABLE 2
%%%%%%%%%%%%%%%%%%%%%%%%%%%%%%%%%
\begin{deluxetable*}{lccccccc}
\tabletypesize{\scriptsize}
%\tabletypesize{\tiny}
\setlength{\tabcolsep}{0.05in}
%\rotate
\tablecaption{A Summary of the TRGB Distance Estimates to SN Ia Host Galaxies}
\tablewidth{0pt}
\tablehead{ \colhead{Target} & \colhead{Region} & \multicolumn{2}{c}{TRGB Magnitude} & Offset$^a$ &  \colhead{$(m-M)_0^b$} & \colhead{$(m-M)_0^b$} & \colhead{$(m-M)_0^b$}  \\
&  & F814W & $QT$ &  & NGC 4258 scale & LMC scale & NGC 4258 + LMC  
}
\startdata
\multicolumn{7}{l}{Blue $I$ calibration} \\

M101 & Entire field &$25.162\pm0.035$ & ... &0.003& $29.185\pm0.035$ & $29.119\pm0.035$ & $29.145\pm0.035$  \\
M66  & $R_{GC}\geq4\farcm3^c$ &$26.232\pm0.087$ &...&0.002&$30.212\pm0.087$ & $30.146\pm0.087$ & $30.175\pm0.087$  \\

M96  & $R_{GC}\geq3\farcm5^c$ &$26.247\pm0.064$ &...&--0.002&$30.237\pm0.064$ & $30.171\pm0.064$ & $30.189\pm0.064$  \\
NGC 4038/39    & Lowest sky$^d$ &$27.696\pm0.049$ &...&--0.002&$31.663\pm0.049$ & $31.597\pm0.049$ & $31.660\pm0.049$  \\
NGC 5584 	   & Lowest sky$^d$ &$27.826\pm0.060$ &...&--0.002&$31.803\pm0.060$ & $31.737\pm0.060$ & $31.784\pm0.060$  \\

NGC 3021 	   & $SMA\geq1\farcm5$ &$28.179\pm0.049$ &...&0.010&$32.198\pm0.049$ & $32.132\pm0.049$ & $32.148\pm0.049$  \\
NGC 3370 	   & $SMA\geq2\farcm0$ &$28.252\pm0.035$ &...&0.018&$32.253\pm0.035$ & $32.187\pm0.035$ & $32.243\pm0.035$  \\
NGC 1309 	   & $SMA\geq2\farcm0$ &$28.457\pm0.044$ &...&0.048&$32.474\pm0.044$ & $32.408\pm0.044$ & $32.469\pm0.044$  \\
\hline
\multicolumn{7}{l}{$QT$ calibration} \\
M101 & Entire field &...&$25.142\pm0.022$&0.003& $29.166\pm0.022$ & $29.147\pm0.022$ & $29.160\pm0.022$  \\
M66  & $R_{GC}\geq4\farcm3^c$ &...&$26.144\pm0.038$&0.025&$30.186\pm0.038$ & $30.167\pm0.038$ & $30.180\pm0.038$  \\
M96  & $R_{GC}\geq3\farcm5^c$ &...&$26.199\pm0.054$&0.013&$30.229\pm0.054$ & $30.210\pm0.054$ & $30.223\pm0.054$  \\
NGC 4038/39    & Lowest sky$^d$ &...&$27.647\pm0.037$&0.015&$31.683\pm0.037$ & $31.664\pm0.037$ & $31.677\pm0.037$  \\

NGC 5584 	   & Lowest sky$^d$ &...&$27.725\pm0.049$ &0.006&$31.757\pm0.049$ & $31.738\pm0.049$ & $31.751\pm0.049$  \\

NGC 3021 	   & $SMA\geq1\farcm5$ &...&$28.146\pm0.033$ &0.021&$32.184\pm0.033$ & $32.165\pm0.033$ & $32.178\pm0.033$  \\
NGC 3370 	   & $SMA\geq2\farcm0$ &...&$28.212\pm0.041$ &0.030&$32.259\pm0.041$ & $32.239\pm0.041$ & $32.253\pm0.041$  \\
NGC 1309 	   & $SMA\geq2\farcm0$ &...&$28.398\pm0.040$ &0.062&$32.477\pm0.040$ & $32.457\pm0.040$ & $32.471\pm0.040$  \\
\hline
\enddata

\label{tab_TRGB}
\tablenotetext{a}{Denotes F814W and the $QT$ magnitude offsets measured from the artificial star experiments.}
\tablenotetext{b}{Quoted errors are random errors only. Systematic errors are estimated to be about 0.068 mag for the NGC 4258 scale, 0.104 mag for the LMC scale, and 0.057 mag for both NGC 4258 and the LMC scales \citep{jan16}.}
\tablenotetext{c}{Indicates hatched regions shown in Figure 1 of \citet{lee13}.}
\tablenotetext{d}{Indicates hatched regions shown in Figure 1 of \citet{jan15}.}

\end{deluxetable*}

Mean TRGB magnitudes and associated uncertainties were measured by running 10,000 simulations of bootstrap resampling as done for our previous studies \citep{lee12, lee13,jan15}.
In each simulation, we made a new sample of RGB stars by sampling a half size of stars randomly from the original RGB sample.
We obtained TRGB magnitudes from the new samples as done for the original sample.
We generated a histogram of the TRGB magnitudes obtained from the simulations and fitted a Gaussian function 
to the histogram, quoting the gaussian mean as a mean TRGB magnitude and the gaussian width as 1$\sigma$ uncertainty.
Measured TRGB values are: F814W$_{TRGB}=28.179\pm0.049, 28.252\pm0.035$ and $28.457\pm0.044$ mag for NGC 3021, NGC 3370, and NGC 1309, respectively. 
Similarly, $QT_{RGB}=28.146\pm0.033$ mag for NGC 3021, $28.212\pm0.041$ mag for NGC 3370, and $28.398\pm0.040$ mag for NGC 1309 were obtained. 
Derived TRGB magnitudes from the $QT$ luminosity functions are systematically brighter than those from $F814W$ luminosity functions, because the $QT$ system uses foreground extinction corrected F814W magnitude (F814W$_0$).

We detected %a 
small photometric offsets from the artificial star experiments as described in Section 2.
The F814W and the $QT$ magnitude offsets at the TRGB levels of three galaxies are measured to be $\Delta F814W_{TRGB}$ $(\Delta QT_{RGB}) =$ 0.010 (0.021) mag for NGC 3021, 0.018 (0.030) mag for NGC 3370, and 0.048 (0.062) mag for NGC 1309. We corrected these magnitude offsets in the distance estimation. 
The F555W -- F814W color offsets we measured at the TRGB levels are ranging from $\Delta(F555W-F814W)=$ 0.048 mag (NGC 3370) to 0.104 mag (NGC 1309). We assume that these small color offsets will not significantly effect the TRGB measurements. %, which are based on $F814W$ magnitude or the $QT$ magnitude. 
For example, a color offset of $\pm$0.1 mag at the mean TRGB color (= F555W -- F814W $\sim 1.8$) changes the $QT$ magnitude %of only 
only by $\pm$0.003 mag. Thus we ignored the color offsets in the following analysis.

%		TRGB	delI	delCol	TRGB	deoQT	deoCol
%N3021	28.179	0.010	0.060	28.146	0.021	0.057
%N3370	28.252	0.018	0.048	28.212	0.030	0.049
%M1309	28.457	0.048	0.100	28.398	0.062	0.104

Distance moduli to target galaxies were derived by applying the recent TRGB calibration presented in our previous study, Paper IV \citep{jan16}.
Two TRGB calibrations are available: the blue $I$ calibration and the $QT$ calibration.
We first apply the blue $I$ calibration, which uses the traditional  $I$ (F814W) magnitude %with
for  blue RGB stars with $F555W-F814W \leq$ 2.1, where the color-dependence of the TRGB is not significant.
There are three absolute TRGB zero-points based on the combination of 
two distance anchors, NGC 4258 and the LMC, to which precise geometric distances are known \citep{hum13, rie16, pie13}. 
Under the photometic system used in this study, the F555W and F814W band combination of ACS/WFC, absolute magnitudes of the TRGB are
$M_{F814W,TRGB}=-4.030\pm0.068$ mag for the NGC 4258 scale, and $M_{F814W,TRGB}=-3.964\pm0.106$ mag for the LMC scale.
A weighted mean of these two is $M_{F814W,TRGB}=-4.008\pm0.057$ mag.
If we adopt this value, we obtain distance moduli of 
%Combining the $F814W$-band apparent magnitude with the absolute magnitude of the TRGB based on both NGC 4258 and the LMC scales, we obtain distance moduli:
$(m-M)_0=32.144\pm0.049$ for NGC 3021, $(m-M)_0=32.243\pm0.035$ for NGC 3370, and $(m-M)_0=32.469\pm0.044$ for NGC 1309. 
Systematic uncertainties are estimated to be 0.057 mag. 

We also apply the $QT$ calibration that uses the $QT$ magnitude with a wider color range of RGB stars so that more RGB stars can be %applied to
used for the TRGB analysis.
The zero-points for the $QT$ calibration are $M_{QT,TRGB}=-4.017\pm0.067$ mag for the NGC 4258 scale, $M_{QT,TRGB}=-3.998\pm0.101$ mag for the LMC scale, and $M_{QT,TRGB}=-4.011\pm0.056$ mag for both NGC 4258 and the LMC scales. 
Adopting the most accurate calibration based on both distance anchors, we obtain $(m-M)_0=32.178\pm0.033$ for NGC 3021, $(m-M)_0=32.253\pm0.041$ for NGC 3370, and $(m-M)_0=32.471\pm0.040$ for NGC 1309. Systematic uncertainties are estimated to be 0.056 mag. 
{\color{red}{{\bf Table \ref{tab_TRGB}}}} lists a summary of distance estimates based on the revised TRGB calibration for the target galaxies in this study.  We also included five other galaxies that were covered in our previous studies \citep{lee12,lee13,jan15}.

\section{Discussion}

\begin{figure*}
\centering
\includegraphics[scale=0.772]{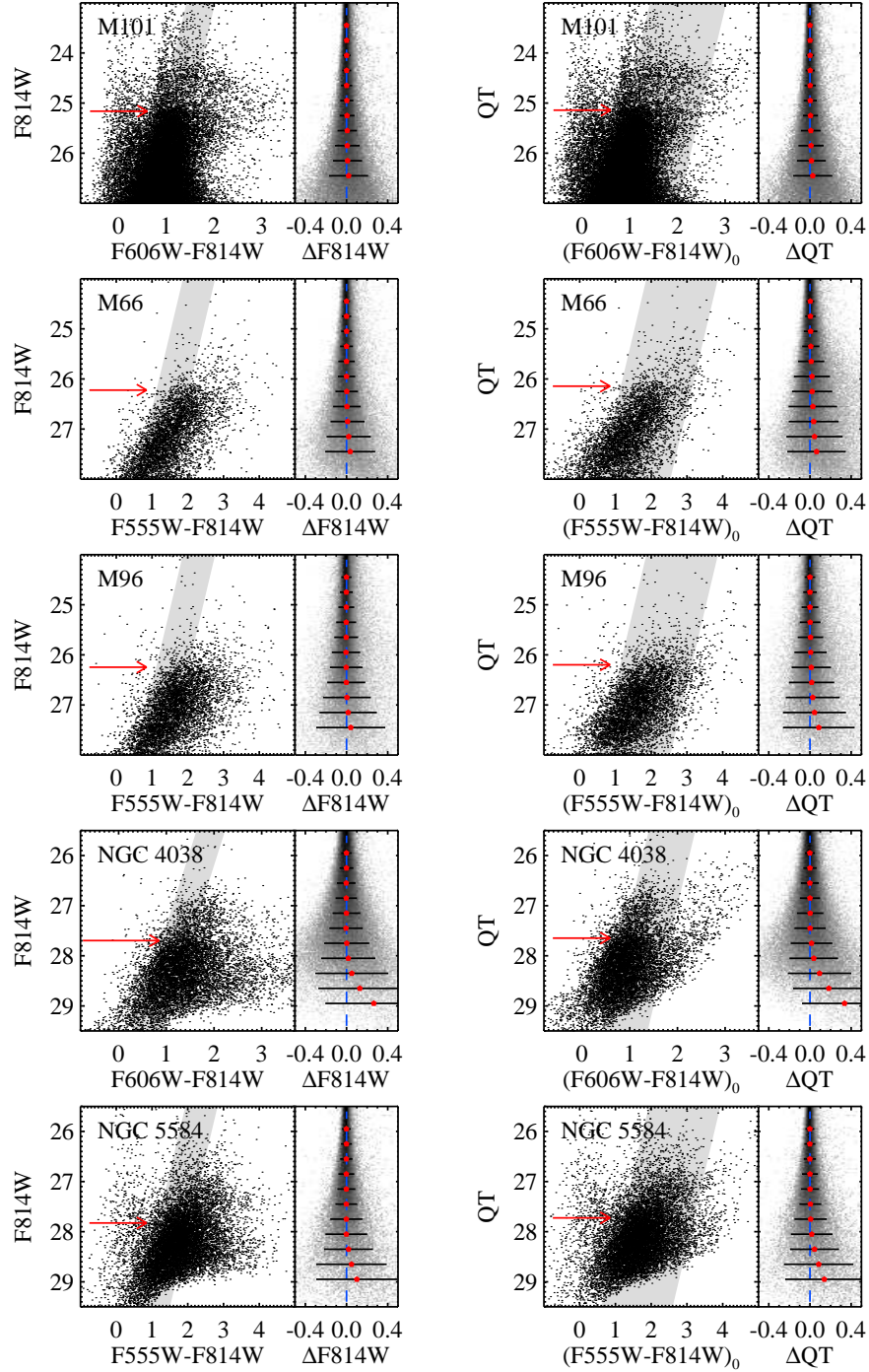} %white.eps}
\caption{
Left windows in each panel: F814W (left panels) and the $QT$ (right panels) CMDs of resolved stars 
in five SN Ia host galaxies used in our previous studies \citep{lee12,lee13,jan15}. 
The shaded region in each panel represents the selection criteria for RGB stars. 
Estimated TRGB magnitudes are marked by red arrows.
Right windows in each panel: results of artificial star experiments in F814W (left panels) and in the $QT$ magnitudes (right panels). Mean offsets with standard deviations in each magnitude bin are marked by red dots with horizontal lines.
}
\label{fig_cmd2}
\end{figure*}

\subsection{Updating the TRGB Distances in Our Previous Studies}

In Papers I \citep{lee12}, II  \citep{lee13}, and III  \citep{jan15}, we presented the TRGB distances to five SN Ia host galaxies: M101 (Paper I), M66 and M96 (Paper II), and NGC 4038/39 and NGC 5584 (Paper III). 
These results were based on the old TRGB calibration \citep{riz07}. We update these results using the revised TRGB calibration introduced in Paper IV. % \citep{jan16}.
%These results are still valid. However, can be updated in this study with the revised TRGB calibration introduced in Paper IV \citep{jan16}.
%We reanalyzed the drizzled images (indicated by \_drc.fits) of five galaxies used in Papers I, II, and III to derive PSF magnitude magnitudes using DAOPHOT/ALLFRAME \citep{ste87, ste94}. 
We carried out PSF photometry on drizzled images (indicated by \_drc.fits) of the five galaxies 
%used in Papers I, II, and III 
using DAOPHOT/ALLFRAME \citep{ste87, ste94}. 
We used the same HST data set as analyzed in our previous studies, except for M101.
In the case of M101, we used new HST/ACS data of a halo field (centered at ($\alpha$, $\delta$)=($14^h03^m47.21^s, 54^\circ14\arcmin25\farcs6$)) %covering the halo of M101 has 
released recently (PID=13364).
Exposure times are 1100s for F606W and 1400s for F814W, longer than those used in Paper I ($\rm F555W=720s\sim1080s$ and $\rm F814W=720s\sim1080s$). 
%We used these deeper image data for M101 in the TRGB analysis.

\begin{figure*}
\centering
\includegraphics[scale=0.7]{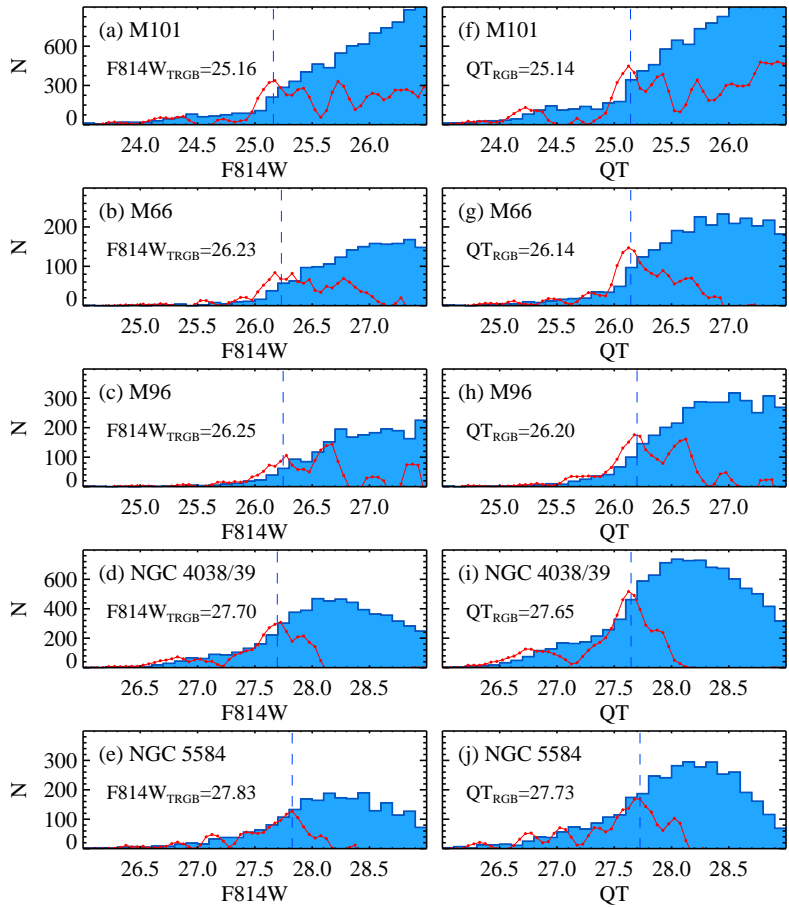} %white.eps}
\caption{
 (a)-(e) F814W magnitude luminosity functions of RGB stars in the selected regions of five SN Ia host galaxies. Edge detection responses and estimated TRGB magnitudes are indicated by red solid and blue dashed lines, respectively.
(f)-(j) Same as (a)-(e) but for the $QT$ magnitude.
}
\label{fig_lf2}
\end{figure*}

{\color{red}{\bf Figure \ref{fig_cmd2}}} displays the CMDs with $F814W$ and the $QT$ magnitudes of the resolved stars in the outer regions of M101, M66, M96, NGC 4038/39 and NGC 5584.
% applied to the $QT$ magnitude.
All of the CMDs show dominant RGB populations, as well as asymptotic giant branch and young main-sequence populations. 
We carried out artificial star experiments to these five galaxies as done for the three galaxies in Section 2, plotting the results in the right side of each CMD. We confirmed that the F814W and the $QT$ magnitude offsets at the expected TRGB level are as small as 0.02 mag. We corrected these offsets in the TRGB distance estimation. %and corrected these offsets in the TRGB distance estimation.

We selected the stars in the shaded regions of CMDs in {\color{red}{\bf Figure \ref{fig_cmd2}}}, producing a sample of RGB stars in each galaxy. 
Then we plotted their luminosity functions and corresponding edge detection responses in {\color{red}{\bf Figure \ref{fig_lf2}}}. 
The luminosity functions of these galaxies show sudden jumps at F814W $\sim QT\sim25.2$ (M101), 26.2 (M66 and M96), 27.7 (NGC 4038/39), and 27.8 (NGC 5584), corresponding to the TRGBs.
It is noted that the edge detection responses for the $QT$ luminosity functions show stronger and more narrow peaks at the TRGB than those for the F814W luminosity functions.
Mean TRGB magnitudes and corresponding errors were measured using the bootstrap resampling 
method as applied to the three galaxies described in Section 3.2.
%We measured the mean TRGB magnitude and their uncertainty using the edge detection algorithm as described in Section 3.2. 
Derived values of the TRGB magnitudes %$QT_{\rm RGB}$ magnitudes, 
are: 
F814W$_{TRGB} = 25.162\pm0.035$ for M101
$26.232\pm0.087$ for M66, 
$26.247\pm0.064$ for M96, 
$27.696\pm0.049$ for NGC 4038/39, and 
$27.826\pm0.060$ for NGC 5584. 
Similarly, we obtain the TRGB magnitudes from the $QT$ luminosity functions:
$QT_{\rm RGB}=25.142\pm0.022$ for M101, 
$26.144\pm0.038$ for M66, 
$26.199\pm0.054$ for M96, 
$27.647\pm0.037$ for NGC 4038/39, and 
$27.725\pm0.049$ for NGC 5584. 
Here all errors are random errors.
The errors from the $QT$ luminosity functions are, on average, 32\% smaller than those of the $F814W$ luminosity functions.

The values of the TRGB magnitudes in the F814W band of these galaxies show excellent agreement ($\Delta (m-M)_0 \lesssim 0.05$) with those in our previous papers, except for M101.
%{\color{red}\bf 
In the case of M101, the new estimate of the TRGB magnitude, F814W$_{TRGB} = 25.162\pm0.035$ is, 0.118 mag brighter than the value in \citet{lee12}, $I_{TRGB}=25.28\pm0.01$.
Main causes for this difference can be explained as follows.
First is the difference in the values of the aperture correction for photometry. 
We adopted a value of 0.098 mag for the aperture correction from $0\farcs5$ to nominal infinity provided by the STScI in this study, whereas \citet{lee12} used an aperture correction of 0.04 mag, which was derived from the data of M101. 
The latter value of aperture correction based on a small number of bright stars in M101 %was 
turned out to be an underestimation.
Second is the difference in the images of M101 used for analysis. We used the images corrected for charge transfer efficiency in this study, while \citet{lee12} used the images without this correction, which were the only available images at that time.

The TRGB distances to these galaxies were obtained by applying the TRGB calibrations, the blue $I$ and the $QT$ calibrations, given in Paper IV. %\citet{jan16}. 
A summary of the TRGB distances of these five galaxies is included in {\color{red}{\bf Table \ref{tab_TRGB}}}.
The distance moduli to the eight SN Ia host galaxies from the two TRGB calibrations show good agreement. 
A weighted mean of distance modulus differences between two calibrations is only $\Delta(m-M)_0=0.012\pm0.021$ mag, when both NGC 4258 and the LMC are used for the TRGB zero-point. 
We use the distance estimates from the $QT$ calibration based on  two distance anchors, NGC 4258 and the LMC, in the following analysis.

%{\color{red} \bf TO REMOVE The difference due to different calibrations in \citet{riz07} and \citet{jan16} is only --0.025!!!.}
%TO ADD MORE????....}
%\clearpage

%%%%%%%%%%%%%%%%%%%%%%%%%%%%%%%%%%%%%%%%%%%
%% TABLE 3
%%%%%%%%%%%%%%%%%%%%%%%%%%%%%%%%%
\begin{deluxetable*}{lcccrr}
\tabletypesize{\footnotesize}
%\tabletypesize{\scriptsize}
%\tabletypesize{\tiny}
\setlength{\tabcolsep}{0.05in}
%\rotate
\tablecaption{A Comparison of Distance Estimates to SN Ia Host Galaxies}
\tablewidth{0pt}
%\tablehead{ \colhead{Target} & \colhead{This study} & \colhead{\citet{rie11}} & \colhead{\citet{rie16}}   \\
\tablehead{ \colhead{Target} & \colhead{This study$^a$} & \colhead{R11$^b$} & \colhead{R16$^b$} & This study -- R11 & This study -- R16 \\
& (TRGB) & (Cepheids) & (Cepheids) 
}
\startdata
M101 		& $29.160\pm0.022$	& ...			& $29.135\pm0.045$	& ...			 &$0.025\pm0.050$\\
NGC 4038/39 & $31.677\pm0.037$  & $31.66\pm0.08$& $31.290\pm0.112$	& $0.017\pm0.088$&$0.387\pm0.118$\\
NGC 5584 	& $31.751\pm0.049$  & $31.72\pm0.07$& $31.786\pm0.046$	& $0.031\pm0.085$&$-0.035\pm0.067$\\
NGC 3021 	& $32.178\pm0.033$  & $32.27\pm0.08$& $32.498\pm0.090$	&$-0.092\pm0.087$&$-0.320\pm0.096$\\
NGC 3370 	& $32.253\pm0.041$  & $32.13\pm0.07$& $32.072\pm0.049$	& $0.123\pm0.081$&$0.181\pm0.064$\\
NGC 1309 	& $32.471\pm0.040$  & $32.59\pm0.09$& $32.523\pm0.055$	&$-0.119\pm0.098$&$-0.052\pm0.068$\\
\hline
\multicolumn{4}{l}{Weighted mean of all}            & $0.001\pm0.039^c$& $0.022\pm0.028^d$ \\
\multicolumn{4}{l}{Weighted mean of all excluding NGC 4038/39}            & $-0.003\pm0.044^e$& $0.000\pm0.029^f$
\enddata

\label{tab_comparison}
\tablenotetext{a}{Based on two distance anchors: NGC 4258 and the LMC.}
\tablenotetext{b}{Based on three distance anchors: NGC 4258, the LMC, and the Milky Way. R11 and R16 denote \citet{rie11} and \citet{rie16}, respectively.}
\tablenotetext{c}{Standard deviation of 0.098 mag.}
\tablenotetext{d}{Standard deviation of 0.235 mag.}
\tablenotetext{e}{Standard deviation of 0.112 mag.}
\tablenotetext{f}{Standard deviation of 0.187 mag.}
\end{deluxetable*}

\subsection{Comparison with the SH0ES Project}

%{\color{red}\bf TO CHECK the numbers below, according to Adam's email}

The SH0ES project has been working on the calibration of SN Ia to measure an accurate value of the Hubble constant as well as other cosmological parameters \citep{rie05, rie09a, rie09b, rie11, rie16}.
They used Cepheid variables, a Population I indicator, for deriving distances to SN Ia host galaxies.
\citet{rie11,rie16} considered four nearby distance anchors for the calibration of Cepheids:
NGC 4258, the LMC, the Milky Way, and M31.
They provided Cepheid distance estimates to SN Ia host galaxies based on three anchors (NGC 4258, the LMC, and the Milky Way) in Table 3 of \citet{rie11} and in Table 5 of \citet{rie16}. %(Riess, private communication on June 25, 2016).
Among 19 and 8 SN Ia host galaxies used in \citet{rie16} and \citet{rie11}, 6
(M101, NGC 4038/39, NGC 5584, NGC 3021, NGC 3370, and NGC 1309) 
and 5 galaxies (the above excluding M101), respectively, are common with those in this study.
Thus, we compare distance estimates
of these galaxies derived in this study with those in \citet{rie11, rie16}, %{\color{red}\bf TO ADD  
summarizing the results in {\color{red}{\bf Table \ref{tab_comparison}}}. % (TRGB, Cepheid(R11) and Cepheid(R16) distances)???}

%Thus, it is possible to compare the distance estimates directly between derived from this study and those from \citet{rie11,rie16}.

\begin{figure}
\centering
\includegraphics[scale=0.9]{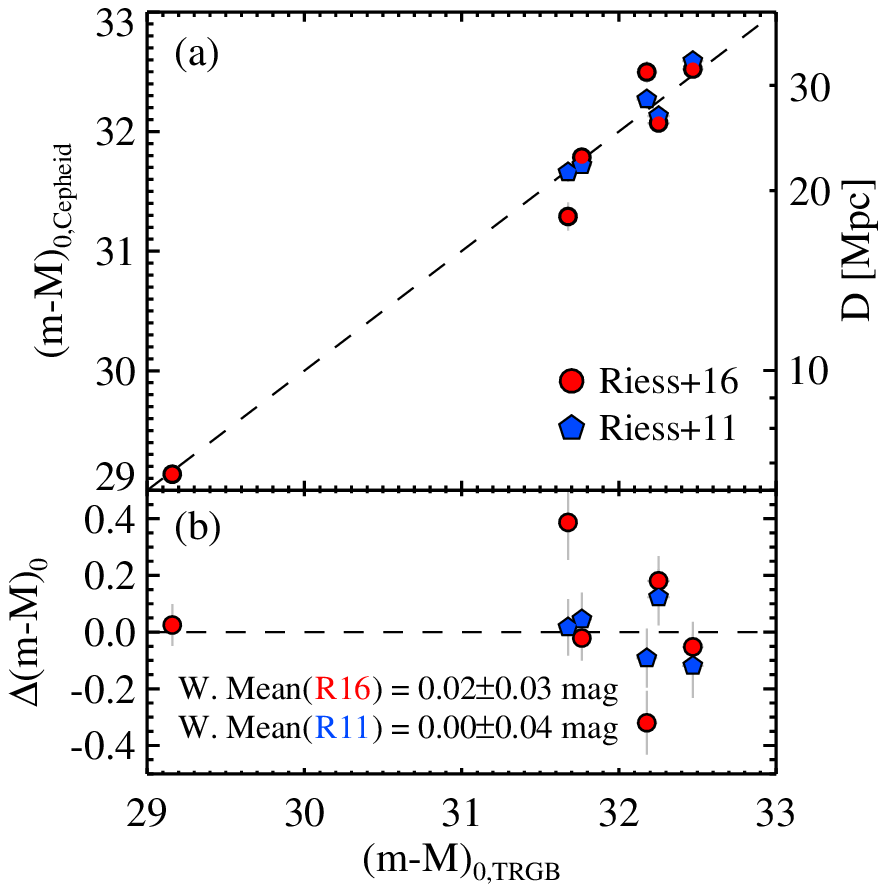} %white.eps}
\caption{ (a) Comparison of distance estimates to SN Ia host galaxies derived in this study based on the TRGB with previous studies based on Cepheids: \citet{rie16} (circles) and \citet{rie11} (pentagons). %, 
%\citet{san06} (triangles), and \citet{fre01} (squares)).
(b) Differences ($\Delta(m-M)_0=(m-M)_{0,{\rm TRGB}}-(m-M)_{0,{\rm Cepheids}}$) between this study and the previous studies.
Weighed mean values of the difference between two studies are labeled.
Note that one outlier with $\Delta(m-M)_0\approx 0.4$ is NGC 4038/39. 
%(b) Distance modulus differences ($\Delta(m-M)_0=$) between this study and previous studies.
%Residuals between the TRGB distance estimates and those from the Cepheids.
%We marked a weighted mean of residual
%(a) A comparison of the TRGB distances in this study with previous studies.
}
\label{fig_compare1}
\end{figure}

{\color{red}{\bf Figure \ref{fig_compare1}(a)}} displays a comparison of the TRGB distance estimates to SN Ia host galaxies derived in this study with Cepheid distance estimates to 6 galaxies in \citet{rie16} (circles) and 5 galaxies in \citet{rie11} (pentagons).
Differences of distance estimates, $\Delta(m-M)_0 = (m-M)_{0,{\rm TRGB}} - (m-M)_{0,\rm Cepheids}$, are shown in {\color{red}{\bf Figure \ref{fig_compare1}(b)}}.
\citet{rie11} adopted a megamaser distance to NGC 4258 of $(m-M)_0=29.31\pm0.05$, which is slightly smaller than the value used in  \citet{rie16} and this study, $(m-M)_0=29.387\pm0.057$.
\citet{rie16} slightly improved the estimate of NGC 4258 distance given by \citet{hum13} and presented an updated value.
%{\color{red}\bf TO REMOVE! We corrected the distance estimates in \citet{rie11}, considering this difference, 0.077 mag.}
However, the distance values in Table 3 of \citet{rie11} and Table 5 of \citet{rie16} are based on the combination of three anchors (NGC 4258, the LMC, and the MW) so the effect of the change for NGC 4258 alone is not as large as 0.077 mag.
\citet{rie16} presented a comparison of their results with those in \citet{rie11} for the  seven common galaxies (except for NGC 4038/39): the mean difference ($\Delta (m-M)_0=$ \citet{rie16} minus \citet{rie11}) is 0.01 mag with a dispersion of 0.12 mag. They noted, however, that the distance difference for NGC 4038/39 (antennae galaxies) hosting SN 2007sr  is as large as $\Delta(m-M)_0 = -0.37$. This large difference is due to removing 10 Cepheids with ultra-long periods (ULP, $P>100$ days) in the sample of Cepheids for NGC 4038/39 during the new analysis of \citet{rie16}.  
 
 %{\color{red}{
%e zeropoint difference of NGC 4258, 0.077 mag, in 
The TRGB distance estimates in this study and the Cepheid distance estimates in \citet{rie11} and \citet{rie16} show 
excellent agreement.
%in general good agreement.
One outlier seen at $\Delta(m-M)_0=0.387\pm0.118$ of {\color{red}{\bf Figure \ref{fig_compare1}(b)}} is NGC 4038/39.
%It is noted that the distance estimates to NGC 4038/39 from \citet{rie11} ($(m-M)_0=31.378\pm0.118$) and \citet{rie16}
% ($(m-M)_0=31.737\pm0.080$) show a significant difference, $0.359\pm0.142$ mag. 
% What causes this difference is not known.%( in \citet{rie11} and  in \citet{rie16}).
A weighted mean of distance difference 
for the five galaxies 
between this study and \citet{rie11} is very small: $\Delta(m-M)_0=0.001\pm0.039$ mag with a standard deviation of 0.098 mag.
A comparison with \citet{rie16} for the six overlapping galaxies yields a similar offset, $\Delta(m-M)_0=0.022\pm0.028$ mag, but its standard deviation is larger than that from \citet{rie11}.
%, showing a weak tension.
If we exclude NGC 4038/39 in the comparison, the mean difference between this study and \citet{rie16} becomes zero, $\Delta(m-M)_0=-0.000\pm0.029$ mag.
%{\color{red}\bf TO CHECK with R16 and REMOVE ?
%If we compare all eight common galaxies used in \citet{rie11} and \citet{rie16}, the mean difference is estimated to be 
%$\Delta(m-M)_0=0.091\pm0.036$ mag with a standard deviation of 0.150 mag. 
%If NGC 4038/39 is excluded, the mean difference will be $\Delta(m-M)_0=0.073\pm0.037$ mag with a standard deviation of 0.117 mag.}

%TO REVISE BELOW LATER
%{\color{red}\bf
%[Revised by In Sung] 
In summary, our TRGB distance estimates are in good agreement with the Cepheid distance estimates in \citet{rie11, rie16}.
A better agreement is seen with those in \citet{rie16}, if we exclude NGC 4038/39.

%%%%%%%%%%%%%%%%%%%%%%%%%%%%%%%%%%%%%%%%%%%
%% TABLE 4
%%%%%%%%%%%%%%%%%%%%%%%%%%%%%%%%%
\begin{deluxetable*}{lccccccccc}
\tabletypesize{\scriptsize}
%\tabletypesize{\tiny}
\setlength{\tabcolsep}{0.05in}
%\rotate
\tablecaption{A Summary of Optical Luminosity Calibration of SNe Ia}
\tablewidth{0pt}
\tablehead{ \colhead{Galaxy} & \colhead{SN Ia} &  \colhead{Filters} & %\colhead{$A_V^{\rm a}$} & \colhead{$m_V^{0 \rm a}+5a_V^{\rm b}$} & \colhead{$m_V^{0}+5a_V^{\rm b}$} & $m_B^{0a}+5a_B^{\rm b}$& \colhead{$(m-M)_0$}&  \colhead{$M_V^{0}$} & \colhead{$M_B^0$}\\
\colhead{$A_V$} & \colhead{$m_V^{0}+5a_V$} & \colhead{$m_V^{0}+5a_V$} & $m_B^{0}+5a_B$& \colhead{$(m-M)_0$}&  \colhead{$M_V^{0}$} & \colhead{$M_B^0$}\\
&  &  &  & This study &  Literature & From R16& TRGB & R11 calibration & R16 calibration
}
\startdata
M101& SN 2011fe & UBVRI & 0.157 &  13.400 &...& 13.310 & $29.160\pm0.022$ & $-19.255\pm0.143$ & $-19.414\pm0.120$ \\
M66 & SN 1989B  & UBVRI & 1.242 & 14.076 & $14.021^{\rm a}$ & ... & $30.180\pm0.038$ & $-19.589\pm0.191$ & ...\\
M96 & SN 1998bu & UBVRI & 1.025 & 14.278 & $14.263^{\rm a}$ & ... & $30.223\pm0.054$ & $-19.430\pm0.182$ & ...\\
N4038/39   & SN 2007sr & uBVri & 0.349 & 15.889 & $15.901^{\rm b}$ & 15.795 & $31.677\pm0.037$ & $-19.273\pm0.149$ & $-19.445\pm0.125$\\
N5584 	   & SN 2007af & uBVri & 0.346 & 16.295 & $16.274^{\rm b}$ & 16.264 & $31.751\pm0.049$ & $-18.941\pm0.152$ & $-19.050\pm0.130$\\
N3021 	   & SN 1995al & BVRI  & 0.213 & 16.630 & $16.699^{\rm b}$ & 16.526 & $32.178\pm0.033$ & $-19.033\pm0.145$ & $-19.215\pm0.123$\\
N3370 	   & SN 1994ae & UBVRI & 0.045 & 16.556 & $16.545^{\rm b}$ & 16.474 & $32.253\pm0.041$ & $-19.182\pm0.146$ & $-19.342\pm0.122$\\
N1309 	   & SN 2002fk & BVRI  & 0.072 & 16.774 & $16.768^{\rm b}$ & 16.755 & $32.471\pm0.040$ & $-19.182\pm0.146$ & $-19.279\pm0.123$\\
\hline
\multicolumn{8}{l}{Weighted mean of eight SNe}            & $-19.209\pm0.055$ & ...\\
\multicolumn{8}{l}{Weighted mean of six low-reddened SNe} & $-19.147\pm0.060$ & $-19.295\pm0.051$\\
%\multicolumn{8}{l}{$H_0$} & $73.05\pm2.04\pm1.55$ & $70.72\pm1.64\pm1.56$\\

\enddata
\label{tab_snia}
%\tablenotetext{{\phantom{a}}}{Note. $A_V$ denotes $V$-band host galaxy extinction
\tablenotetext{a}{Note. $A_V$ denotes $V$-band host galaxy extinction derived from the light curve fitting with the MLCS2k2 code. $m_V^0$ and $m_B^0$ indicate the corrected apparent peak magnitude of SN Ia in $V$ and $B$-band, respectively. $a_V$ ($=0.697\pm0.002$) and $a_B$ ($=0.7127\pm0.0017$) denote the $V$ and $B$-band intercepts of the SN Ia Hubble diagram.}
%\tablenotetext{a}{$A_V$, $m_V^0$ indicate the $V$-band host galaxy extinction and the corrected apparent peak magnitude of SN Ia expressed in \citet{jha07}, respectively.}
%\tablenotetext{b}{$a_V$ ($=0.697\pm0.002$) and $a_B$ ($=0.7127\pm0.0017$) denote the $V$ and $B$-band intercepts of the SN Ia Hubble diagram. Values are obtained from \citet{rie11} for $a_V$ and \citet{rie16} for $a_B$.}
\tablenotetext{a}{Derived from \citet{jha07}.}
\tablenotetext{b}{Derived from \citet{rie11}.}
%\tablenotetext{e}{Quoted errors are quadratic sums of the TRGB distance errors, extinction errors (10\% of the Milky Way and the host galaxy extinction values), and the luminosity dispersion of SNe Ia (0.14 mag \citep{jha07}).}
\end{deluxetable*}

\subsection{A Distance Discrepancy for NGC 4038/39}

The significant difference ($\Delta(m-M)_0({\rm TRGB-Cepheid})=0.387\pm0.118$) for NGC 4038/39 between the TRGB distance in this study
and the Cepheid distance in \citet{rie16} is 
an interesting issue, deserving %to investigate
an investigation of any cause for this discrepancy.
NGC 4038/39 is a pair of interacting galaxies, showing clear features of strong star formation. 
However, the CMD of the resolved stars derived from deep HST images of a region outside arms in NGC 4038/39 shows a prominent RGB as well as a weaker AGB. Photometry of the RGB stars reaches about 1.5 mag below the TRGB so that the detection of the TRGB is considered to be solid \citep{jan15}. 
The TRGB distance for NGC 4038/39 in this study and \citet{jan15} is consistent with the previous estimate based on the TRGB  by \citet{sch08}, $(m-M)_0= 31.51 \pm 0.17$.
%\citet{jan15} describes what causes this small difference between these two values in detail. 
Thus the TRGB distance estimate for this galaxy in this study is considered to be reliable.

On the other hand, two recent studies based on the same data of Cepheids, \citet{rie11} and \citet{rie16}, showed a large discrepancy.
The reason for this difference is related with the presence of ULP Cepheids in the Cepheid sample for NGC 4038/39.
ULP Cepheids have longer periods ($P>80$ days) and are brighter ($-6> M_V >-8$) than classical Cepheids. They are often found in star-forming late-type galaxies with low-metallicity and their period-luminosity relation is not as well-known as classical Cepheids \citep{bir09,fio12,fio13}. 

\citet{rie16} excluded 10 ULP Cepheids %with ultra-long period 
from the sample of Cepheids for NGC 4038/39 used in \citet{rie11} for distance estimation, 
because the phase coverage of these Cepheids is very sparse and because the period-luminosity relation of ULP Cepheids is not as reliable as that of normal Cepheids. % (see also \citet{fio12,fio13}). 
%the poorly constrained properties of the P-L relation for these intrinsically rare objects (Bird et al. 2009; Fiorentino et al. 2012).
From the Cepheid sample without ULP Cepheids, \citet{rie16} obtained a distance modulus for NGC 4038/39, 0.37 mag smaller than the value derived from the full sample of Cepheids that was used in \citet{rie11}. 
\citet{fio13} applied a theoretical scenario for classical Cepheids to the same sample of Cepheids (including ULP Cepheids) in the galaxies as used in \citet{rie11}. They derived a distance to NGC 4038/39 from the full sample of Cepheids, $(m-M)_0= 31.55
\pm 0.06$, which is consistent with the value in \citet{rie11}, but about 0.3 larger than the value given by \citet{rie16}. 
NGC 4038/39 shows the largest fraction of ULP Cepheids in the sample of SN Ia host galaxies used by \citet{rie11,rie16}, which must be strongly related with active star formation activity. 
Considering that most of the Cepheids in Cepheid distance anchors are classical Cepheids, it appears to be conservative to use only classical Cepheids in the case of NGC 4038/39. However, it is difficult to 
understand the large difference between this value and the TRGB distance.
A further study is needed to investigate the effect of the ULP Cepheids in Cepheid distance estimation \citep{fio12}.

%Schweizer et al. (2008) TRGB 31.51 ± 0.17 TRGB = 27.46 ± 0.12
%This Study TRGB 31.67 ± 0.05 ITRGB = 27.67 ± 0.05

%These results indicate that the distance to NGC 4038/39 is closer to the TRGB distance in this study and the Cepheid distance in \citet{rie11}, rather than to the Cepheid distance in \citet{rie16}.}
%It should be explained in the future..?

% weighted mean of eight SNe used both in R11 and R16. mean=0.091, meanerr, 0.036, stdev=0.150

%We added 0.077 mag to the estimates from \citet{rie11} to be consistent with 

%\citet{rie11} adopted $(m-M)_0=29.31\pm0.05$ for the ditance to NGC 4258.

%Residuals of distance estimates between two methods are shown in Figure \ref{fig_compare1} (b).

% Rie09a
% Rie09b
% Rie11 optical pl, F160W luminosity
% Rie16 optical pl, F160W luminosity

\subsection{The Calibration of SNe Ia and the Hubble Constant}

With the TRGB distances to SN Ia host galaxies derived in this study, we calibrate the absolute peak luminosity of SNe Ia and estimate the value of the Hubble constant.
Optical light curves of eight SNe Ia were obtained from various literature: $UBVRI$ photometric data for SN 2011fe from \citet{per13}, for SN 1989B from \citet{wel94}, for SN 1998bu from \citet{jha99}, for SN 1995al from \citet{rie09a}, 
$BVRI$ photometric data for SN 1994ae and SN 2002fk from \citet{rie09a}, and $uBVri$ photometric data for SN 2007sr and SN 2007af from the Carnegie Supernova Program \citep{ham06} (u) and \citet{hic09} (BVri).
We also compiled three $V$-band observations at the pre-maximum of SN 2007sr provided by the All-Sky Automated Survey (ASAS) \citep{poj97} and \citet{rie11}.
All eight SNe Ia were observed with CCD detectors, so that detector dependent uncertainties are estimated to be smaller than those from photographic or photoelectric detectors.

We derived light-curve parameters of each SN Ia using the MLCS2K2 code (version 0.07) \citep{jha07}. % and summarized the results in Table \ref{tab_snia}.
%Light curve parameters of each SN Ia were derived from the MLCS2K2 code (version 0.07) \citep{jha07}, as summarized in Table \ref{tab_snia}.
We set total to selective extinction ratio, $R_V=3.1$ for the Milky Way and $R_V$=2.5 for the SN Ia host galaxies, to be consistent with the values used in \citet{rie11,rie16}. 
The derived values of the parameters, $A_V$ and $m_V^0+5a_V$ (where $a_V=0.697\pm0.002$ \citep{rie11}), are listed in {\color{red}{\bf Table \ref{tab_snia}}}.
We also listed the $m_V^0+5a_V$ values from \citet{jha07} and \citet{rie11} in column six of {\color{red}{\bf Table \ref{tab_snia}}}.
The values of $m_V^0+5a_v$ derived in this study agree well with those from \citet{jha07} and \citet{rie11}, showing a mean difference of 0.004 mag.
%A mean $m_V^0$ difference of seven SNe, excluding SN 2011fe in M101, is only 0.004 mag, which is negligible.
%Obtained light curve parameter values of $A_V$ and $m_V^0$ are listed in Table \ref{tab_snia}.
%Obtained light curve parameter values ($A_V$, and $m_V^0$) agree well with the values listed in \citet{jha07, rie09b,rie11}, within the fitting uncertainty, 0.08 mag \citep{rie11}.
Absolute maximum magnitudes of eight SNe Ia, 
corrected to the fiducial color and luminosity, 
were also derived from the $m_V^0$ values and the TRGB distances derived in this study, as listed in the column nine of {\color{red}{\bf Table \ref{tab_snia}}}.
Quoted uncertainties are quadratic sums of the TRGB distance errors, extinction errors (10\% of the Milky Way and the host galaxy extinction values), and the luminosity dispersion of SNe Ia, 0.14 mag \citep{jha07}.

\begin{figure}
\centering
\includegraphics[scale=0.85]{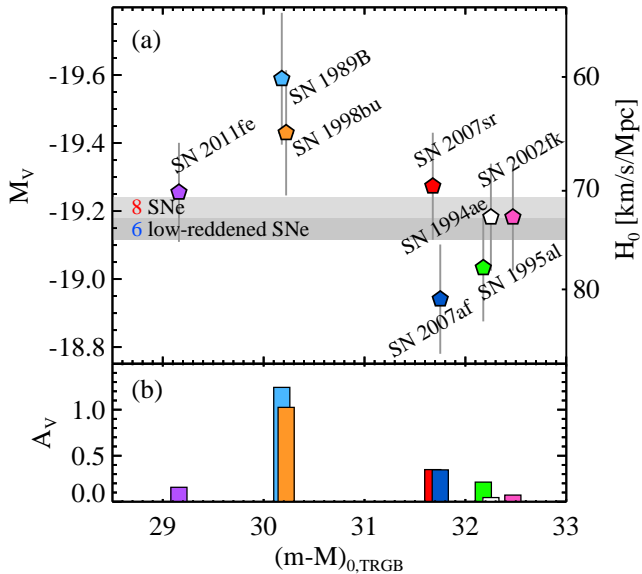} %white.eps}
\caption{(a) Comparison of absolute peak magnitudes of eight SNe Ia with their TRGB distances derived in this study. Two shaded regions indicate weighted means of eight SNe Ia (upper region) and six low-reddened SNe Ia (lower region). 
%We mark the values of the Hubble constant derived from the mean absolute magnitudes of SNe Ia.
(b) Comparison of $V$-band host galaxy extinction values ($A_V$) with the TRGB distances. Note that two SNe, SN 1989B in M66 and SN 1998bu in M96, 
show significantly higher extinctions than those of other six SNe. 
}
\label{fig_compare2}
\end{figure}

{\color{red}{\bf Figure \ref{fig_compare2}}} displays distributions of $V$-band absolute maximum magnitudes ($M_V^0$)  and host galaxy extinction values ($A_V$) of SNe Ia as a function of the TRGB distances derived in this study. 
Absolute magnitudes of SNe Ia range from $M_V^0=-18.941$ (SN 2007af) to $-19.589$ (SN 1989B). 
Host galaxy extinction values are estimated to be small, except for two SNe.
SN 1989B and SN 1998bu show significantly higher extinction values ($A_V>1.0$) than those of other SNe ($A_V<0.4$).
Thus derived absolute magnitudes for these two highly reddened SNe are probably less reliable than the others, although we corrected for their extinctions.
The standard deviation of all eight SNe and six low-reddened SNe are 0.206 mag and 0.131 mag, respectively. 
The former is slightly larger than the intrinsic luminosity dispersion of SNe Ia derived from a large sample of SNe, 0.14 mag \citep{jha07} and 0.128 mag \citep{rie16}. 
On the other hand, the value for the six low-reddened SNe is similar to the intrinsic luminosity dispersion of SNe.
%but the latter is similar.
%The value from eight SNe is slightly larger than the intrinsic luminosity dispersion of SNe, 0.14 mag \citep{jha07} or 0.128 mag \citep{rie16}, but that from the low-reddened SNe is similar.

A weighted mean of the absolute peak magnitude of eight SNe is $M_V^0=-19.209\pm0.055$ mag.
It is 0.089 mag brighter than that from the SH0ES project \citep{rie11,rie16} ($M_V^0=-19.12$ mag).
% and 0.238 mag fainter than those from the Hubble key project \citep{fre01} and the SN HST project \citep{san06} ($M_V=-19.46$ mag).
If we use six low-reddened SNe, then a weighted mean of the absolute peak magnitude would be slightly fainter, $M_V^0=-19.147\pm0.060$, getting closer to the estimate by \citet{rie11,rie16}.
\citet{rie16} provided SALT-II fits of 19 SNe Ia including the 6 low-reddened SNe used in this study. They did not included 2 highly reddened SNe, SN 1989B and SN 1998bu.
The values of $m_B^0+5a_B$ provided by \citet{rie16} are listed in column 7 of {\color{red}{\bf Table \ref{tab_snia}}}.
The $B$-band absolute magnitudes of SNe Ia ($M_B^0$) were derived from the $m_B^0$ values in \citet{rie16} and the TRGB distance estimates in this study, as listed in the last column of {\color{red}{\bf Table \ref{tab_snia}}}.
%We derived the absolute magnitudes of SNe Ia ($M_B^0$) from the $m_B^0$ values in \citet{rie16} and the TRGB distance estimates in this study and listed the values in column seven and ten of {\bf Table \ref{tab_snia}}
%The values of $m_B^0+5a_B$ provided by \citet{rie16} and corresponding absolute magnitudes ($M_B^0$) derived from our TRGB distances are listed in column seven and ten of {\bf Table \ref{tab_snia}},
%respectively.
A weighted mean of the $B$-band absolute magnitudes of the six SNe is $M_B^0=-19.295\pm0.051$ mag. 
It is 0.045 mag brighter than that given in \citet{rie16}, $M_B^0=-19.25$ mag.%$M_B^0=-19.257\pm0.033$ mag.

\citet{rie11, rie16} %presented 
used an equation deriving the value of $H_0$ from the absolute magnitude of SNe Ia:

\begin{displaymath}
log H_0 = \frac{M_x^0 + 5a_x + 25}{5}
\end{displaymath}
\noindent where $M_x^0$ is an absolute magnitude of SNe Ia and $a_x$ is an intercept of the SN Ia Hubble diagram in band-x. 

\citet{rie11} presented a value of $a_V$ ($=0.697\pm0.00201$) that was determined from the MLCS2k2 fits of 140 SNe Ia at $0.023<z<0.10$. 
%A value of $a_V$, $0.697\pm0.00201$, was determined from the MLCS2k2 fits of 140 SNe Ia at $0.023<z<0.10$ in \citet{rie11}.
%From the MLCS2k2 fits of 140 SNe Ia at $0.023<z<0.10$, \citet{rie11} yielded the value, $a_V=0.697\pm0.00201$.
%Similarly, \citet{rie16} presented the value, $a_B=0.7127\pm0.0017$, from the SALT-II fits of 233 SNe Ia at $0.023<z<0.15$.\\
Combining the above equation with the $V$-band absolute magnitudes of SN Ia derived in this study, we derive a value of the Hubble constant: $H_0=71.66\pm1.80 (\rm random)\pm1.88 (systematic)$ \kmsMpc from the eight SNe, and $H_0=73.72\pm2.03 (\rm random)\pm1.94 (\rm systematic)$ \kmsMpc from the six low-reddened SNe.

\begin{figure*}
\centering
\includegraphics[scale=1.0]{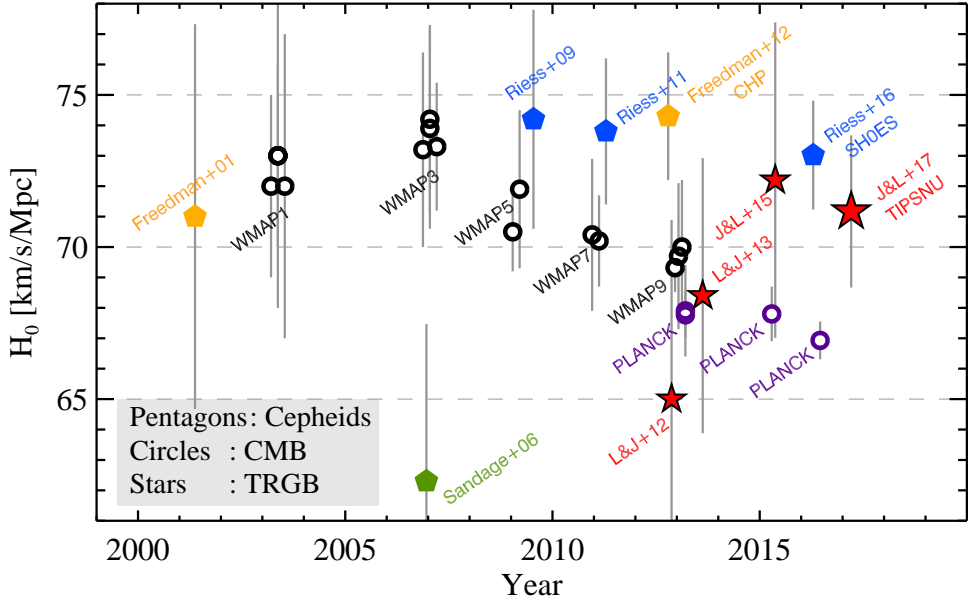} %white.eps}
\caption{Comparison of recent values of the Hubble Constant based on the TRGB calibration of SN Ia (starlets), the Cepheid calibration of SN Ia (pentagons), and the WMAP and Planck CMBR results (circles). Note that our best TRGB estimate in this study is between the 2016 Cepheid value and recent Planck values.
}
\label{fig_H0time}
\end{figure*}

\citet{rie16} provided an updated value,  $a_B=0.7127\pm0.0017$ from the SALT-II fits of 233 SNe at $0.023<z<0.15$.
If we use this value and the sample of six low-reddened SNe, we obtain our best estimate of the Hubble constant, which is accurate to 3.51\% ($\pm2.50$ \kmsMpc) including the systematics: $H_0=71.17\pm1.66 (\rm random)\pm1.87 (systematic)$ \kmsMpc. %, which is our best estimate.
\citet{rie16} also present a value of the Hubble constant, when only two distance anchors (NGC 4258 and the LMC) are used for Cepheid distances:  $H_0=71.61 \pm 1.78$ \kmsMpc. Thus, our best estimate is consistent with this value.
This shows that the TRGB  and the Cepheid produce distance results consistent with each other, as long as the same distance anchors are used.

In {\color{red}{\bf Figure \ref{fig_H0time}}}, we compare  recent values of the Hubble constant based on the TRGB calibration of SN Ia in this study, including \citet{lee12,lee13}, \citet{jan15} the Cepheid calibration of SN Ia \citep{fre01,san06,rie09b,rie11,fre12,rie16}, and the WMAP and PLANCK CMBR results.
% Note that our best TRGB estimate in this study is between the 2016 Cepheid value and the 2015 Planck value.
Our best estimate is between the values from the Cepheid calibrated SNe Ia based on all four distance anchors and those from the CMB analysis. 
The levels of differences between our best estimate and the values from other studies are: $1.70 \sigma$ with \citet{pla15} (PLANCK), $0.92 \sigma$ with \citet{ben13} (WMAP9), $0.56 \sigma$ with \citet{rie16} (SH0ES), $0.88 \sigma$ with \citet{rie11} (SH0ES), and $1.08 \sigma$ with \citet{fre12} (CHP-I).
Thus, our estimate agrees, within $1.7 \sigma$, with the recent estimations based on Cepheid calibrated SNe Ia and the CMB analysis.
%Our estimate shows $1.26 \sigma$, $0.61 \sigma$, $1.02 \sigma$, $1.36\sigma$, $1.58 \sigma$
The Cepheid calibrated results in \citet{rie16} heighten the Hubble tension. On the other hand, our TRGB calibrated results are comfortably located between the Cepheid results and the CMB results, lowering the Hubble tension.

In this study, we used two distance anchors, NGC 4258 and the LMC, while recent Cepheid studies \citep{rie11,rie16} utilized two more anchors,
the Milky Way Cepheids with parallaxes and M31 for which geometric distances of two DEBs are known \citep{rib05,vil10} ($(m-M)_0 = 24.36
\pm0.08$). 
Note that the two anchors used in this study %favor 
lead to a slightly lower value of the Hubble Constant based on Cepheids, than the other two anchors, the Milky Way and M31: $H_0 = 72.2\pm2.51$ for NGC 4258, $H_0 = 72.04\pm2.67$ for the LMC, $H_0 = 76.18\pm2.37$ for the Milky Way, and $H_0 = 74.50\pm3.27$ for M31 \kmsMpc \citep{rie16}. 
Thus it is important to understand any causes for this difference and to reduce this difference in the future. 
We are planning to investigate M31 as another anchor for the TRGB in the near future.
When GAIA astrometry of the Milky Way stars is available, a number of TRGB stars in the Milky Way can be used for absolute calibration of the TRGB magnitudes so they can serve as a distance anchor for the TRGB \citep{deb14,bea16}.

%\citet{rie16} presented an updated value, $a_B=0.7127\pm0.0017$, from the SALT-II fits of 233 SNe Ia at $0.023<z<0.15$.
%Similarly, we obtain the value of the Hubble constant from the $B$-band absolute magnitudes of SNe Ia : $H_0=70.69\pm1.67\pm1.56$ \kmsMpc.

%\citet{rie16} provides SALT2 fits of eight SNe Ia including the six low-reddened SNe used in this study. 

%We compiled the $M_B^0+5a_B$
%Using the $M_B^0+5a_B$ provided by \citep{rie11} and the TRGB distances in this study, 

% Rv = 3.1, 2.5
% 10% extinction error
% 0.14 mag error
% consistent with Riess 11.

% M101 \citet{per13} UBVRI
% M66 \citet{wel94}  UBVRI
% M96 \citet{jha99}  UBVRI
% NGC 4038/39 CSP (u) + \citet{hic09} uBVri
% NGC 5584 CSP (u) + \citet{hic09}    uBVri
% NGC 3021 \citet{rie09a} UBVRI
% NGC 3370 \citet{rie99} BVRI
% NGC 1309 \citet{rie09a} BVRI

%\subsection{Comparison with Previous Distance Estimates}

%\subsection{The Calibration of Near Infrared Maximum Magnitude}
%\subsection{The Calibration of the Hubble Constant}

\section{Summary}

Using deep photometry of the resolved stars in the HST images, we determine  TRGB distances  to three SN Ia host galaxies: 
NGC 3021 hosting SN 1995al, NGC 3370 hosting SN 1994ae, and NGC 1309 hosting SN 2002fk. We combine these results with those in our previous studies and calibrate the luminosity of SN Ia. Then we estimate the value of the Hubble Constant using the SN Ia with the TRGB calibration.
Primary results are as follows.

\begin{itemize}

\item We find a significant number of old RGB stars in the halo regions of the three galaxies, presenting their CMDs.
%SN Ia host galaxies from tphotometry of the combined archival $HST/ACS$ data.
%We obtain the deep photometry of resolved stars in the halos of NGC 3021, NGC 3370, and NGC 1309.
%from the master drizzled images.
%detection of RGB stars

\item Applying the revised TRGB calibration introduced by \citet{jan16} (Paper IV) we measured the TRGB magnitudes, obtaining $QT_{\rm RGB}$=$28.146\pm0.033$ for NGC 3021, $28.212\pm0.041$ for NGC 3370 and $28.398\pm0.040$ for NGC 1309.
From these, we derive TRGB distances: $(m-M)_0=32.178\pm0.033$ for NGC 3021, $32.253\pm0.041$ for NGC 3370, and $32.471\pm0.040$ for NGC 1309.

\item We updated our previous distance estimates to five SN Ia host galaxies (M101, M66, M96, NGC 4038/39 and NGC 5584),
as listed in {\color{red}{\bf Table \ref{tab_TRGB}}},  applying the revised TRGB calibration.

\item We compared our TRGB distance estimates with the Cepheid distance estimates presented in \citet{rie11} and \citet{rie16}, 
 obtaining \\ $\Delta (m-M)_0 ({\rm TRGB - Cepheid})=0.001\pm0.039$ mag for five common galaxies in \citet{rie11}, $0.022\pm0.028$ mag for six common galaxies in \citet{rie16}, and  $0.000\pm0.029$ mag for five galaxies (excluding NGC 4038/39) in \citet{rie16}.

\item Our TRGB distance to NGC 4038/39 is  significantly larger than the Cepheid distance given by \citet{rie16} ($\Delta(m-M)_0({\rm TRGB-Cepheid})=0.387\pm0.118$), while it is similar to the Cepheid distance given by \citet{rie11}. The difference in the Cepheid distance between \citet{rie11} and \citet{rie16} is due to the presence of ULP Cepheids in this galaxy. 
This shows that the Cepheid distance estimation of active star-forming galaxies can be affected significantly by the ULP Cepheids. 

\item From the mean absolute peak magnitude of the six low-reddened SNe Ia and  the recent SN Ia calibration in \citet{rie16}, we derive our best estimate of the Hubble constant: $H_0=71.17\pm1.66 (\rm random)\pm1.87 (systematic)$ \kmsMpc. This  estimate is between the values from the Cepheid calibrated SNe Ia \citep{rie11, fre12, rie16} and those from the CMB analysis \citep{ben13, pla15}. This alleviates the Hubble tension between the Cepheid results and CMB results.

\end{itemize}

\bigskip
The authors are grateful to the anonymous referee for
his/her comments that improved the original manuscript and to 
%The authors are thankful to 
Adam Riess for useful discussions on Cepheid distances to SN Ia host galaxies.
This work was supported by the National Research Foundation of Korea (NRF) grant
funded by the Korea Government (MSIP) (No. 2012R1A4A1028713).
This paper is based on image data obtained from the Multimission Archive at the Space Telescope Science Institute (MAST).

\clearpage

\clearpage

\clearpage

%(a), (b), and (c)) LFs (histograms) and edge detection responses (curved lines) of the RGB stars in NGC 3021 (a), NGC 3370 (b), and NGC 1309 (c). 
%Estimated TRGB magnitudes are marked by vertical dashed lines in each panel.
%((d), (e), and (f)) Same as (a), (b), and (c), except for the $QT$ magnitude.

\clearpage

\end{document}